\documentclass[aps, prx, twocolumn, superscriptaddress, ]{revtex4-1}

\usepackage{amsmath,amssymb,amsfonts,amsbsy}
\usepackage{graphicx}
\usepackage{dcolumn}
\usepackage{bm}
\usepackage{multirow}
\usepackage{mathtools}
\usepackage{array}
\usepackage{hyperref}
\usepackage{color}
\usepackage[normalem]{ulem}

\newcommand{\app}[1]{\hyperref[app:#1]{Appendix~\ref*{app:#1}}}

\newcommand{\ket}[1]{\left|#1\right\rangle}

\begin{document}

\title{Repetitive quantum non-demolition measurement and soft decoding of \\a silicon spin qubit}

\author{Xiao~Xue}
 \thanks{These authors contributed equally to this work}
\affiliation{QuTech and Kavli Institute of Nanosicence, Delft University of Technology, Lorentzweg 1, 2628 CJ Delft, The Netherlands}
\author{Benjamin~D'Anjou}
 \thanks{These authors contributed equally to this work}
\affiliation{Department of Physics, University of Konstanz, D-78457 Konstanz, Germany}
\author{Thomas~F.~Watson}
\affiliation{QuTech and Kavli Institute of Nanosicence, Delft University of Technology, Lorentzweg 1, 2628 CJ Delft, The Netherlands}
\author{Daniel~R.~Ward}
\affiliation{University of Wisconsin-Madison, Madison, WI 53706, USA}
\author{Donald~E.~Savage}
\affiliation{University of Wisconsin-Madison, Madison, WI 53706, USA}
\author{Max~G.~Lagally}
\affiliation{University of Wisconsin-Madison, Madison, WI 53706, USA}
\author{Mark~Friesen}
\affiliation{University of Wisconsin-Madison, Madison, WI 53706, USA}
\author{Susan~N.~Coppersmith}
 \thanks{Present address: School of Physics, University of New South Wales, Sydney NSW 2052, Australia.}
\affiliation{University of Wisconsin-Madison, Madison, WI 53706, USA}
\author{Mark~A.~Eriksson}
\affiliation{University of Wisconsin-Madison, Madison, WI 53706, USA}
\author{William~A.~Coish}
\affiliation{Department of Physics, McGill University, Montreal, Quebec H3A 2T8, Canada}
\author{Lieven~M.~K.~Vandersypen}
 \email{Correspondence author: l.m.k.vandersypen@tudelft.nl}
\affiliation{QuTech and Kavli Institute of Nanosicence, Delft University of Technology, Lorentzweg 1, 2628 CJ Delft, The Netherlands}
\date{\today}

\begin{abstract}

Quantum error correction is of crucial importance for fault-tolerant quantum computers. As an essential step towards the implementation of quantum error-correcting codes, quantum non-demolition (QND) measurements~\cite{braginsky1980quantum, braginsky1996quantum, ralph2006quantum} are needed to efficiently detect the state of a logical qubit without destroying it~\cite{schaetz2005enhanced, hume2007high, pla2013high, jiang2009repetitive, robledo2011high, saira2014entanglement, elder2019high}. Here we implement QND measurements in a Si/SiGe two-qubit system ~\cite{watson2018programmable}, with one qubit serving as the logical qubit and the other serving as the ancilla. Making use of a two-qubit controlled-rotation gate, the state of the logical qubit is mapped onto the ancilla, followed by a destructive readout of the ancilla. Repeating this procedure enhances the logical readout fidelity from $75.5\pm 0.3\%$ to $94.5 \pm 0.2\%$ after 15 ancilla readouts. In addition, we compare the conventional thresholding method with an improved signal processing method called soft decoding that makes use of analog information in the readout signal to better estimate the state of the logical qubit~\cite{d'anjou2014soft}. We demonstrate that soft decoding leads to a significant reduction in the required number of repetitions when the readout errors become limited by Gaussian noise, for instance in the case of readouts with a low signal-to-noise ratio. These results pave the way for the implementation of quantum error correction with spin qubits in silicon.

\end{abstract}

\maketitle

The compatibility of spin qubits with industrial semiconductor technology as well as their relatively small size makes them scalable to large dense arrays~\cite{vandersypen2017interfacing, li2018crossbar} and facilitates the implementation of fault-tolerant quantum computing based on quantum error correction. A key requirement of quantum error correction is the ability to repeatedly measure multiple physical qubits in a quantum non-demolition (QND) way to identify logical errors~\cite{fowler2012surface}. One approach to achieve quantum non-demolition readout of spin qubits is to use a two-qubit gate to map the state of the logical qubit to an ancilla which is then measured. While the readout of the ancilla may be destructive, it leaves the state of the original qubit unperturbed. Consequently, the ancilla may be reinitialized and the logical qubit measurement can be repeated to enhance the signal. Recently, ancilla-based repetitive QND readout has been implemented across several platforms, from trapped ions~\cite{schaetz2005enhanced, hume2007high} to electron-nuclear spin systems~\cite{pla2013high, jiang2009repetitive, robledo2011high} and superconducting qubits~\cite{saira2014entanglement, elder2019high}. In GaAs quantum dots, repeated non-destructive readout of spin states~\cite{meunier2006nondestructive} as well as QND measurement of a spin qubit~\cite{nakajima2019quantum}  have been reported. In the latter experiment, however, the information had to be decoded from the evolution of the ancilla qubit under a two-qubit controlled-phase operation with variable interaction time to overcome the fluctuations of the Overhauser field. This makes the cumulative fidelity only slowly increase with the number of QND readouts. In addition, the binary thresholding of individual measurements in these experiments discards valuable information. Furthermore, future experiments on large arrays of spin qubits will likely rely on gate-based dispersive readout, where it is challenging to achieve a high signal-to-noise ratio (SNR)~\cite{pakkiam2018single, west2019gate, urdampilleta2019gate, zheng2019rapid}. Consequently, both a repetitive QND readout with uniform repetitions and improved decoding methods are highly desirable for quantum error correction. In particular, ``soft decoding" makes full use of the analog information contained in the successive measured detector responses. Such analog information was previously used in trapped ion experiments~\cite{hume2007high} and can lead to a more efficient readout of the logical qubit~\cite{d'anjou2014soft}. 

\begin{figure*}[t] 
\center{\includegraphics[width=1.0\linewidth]{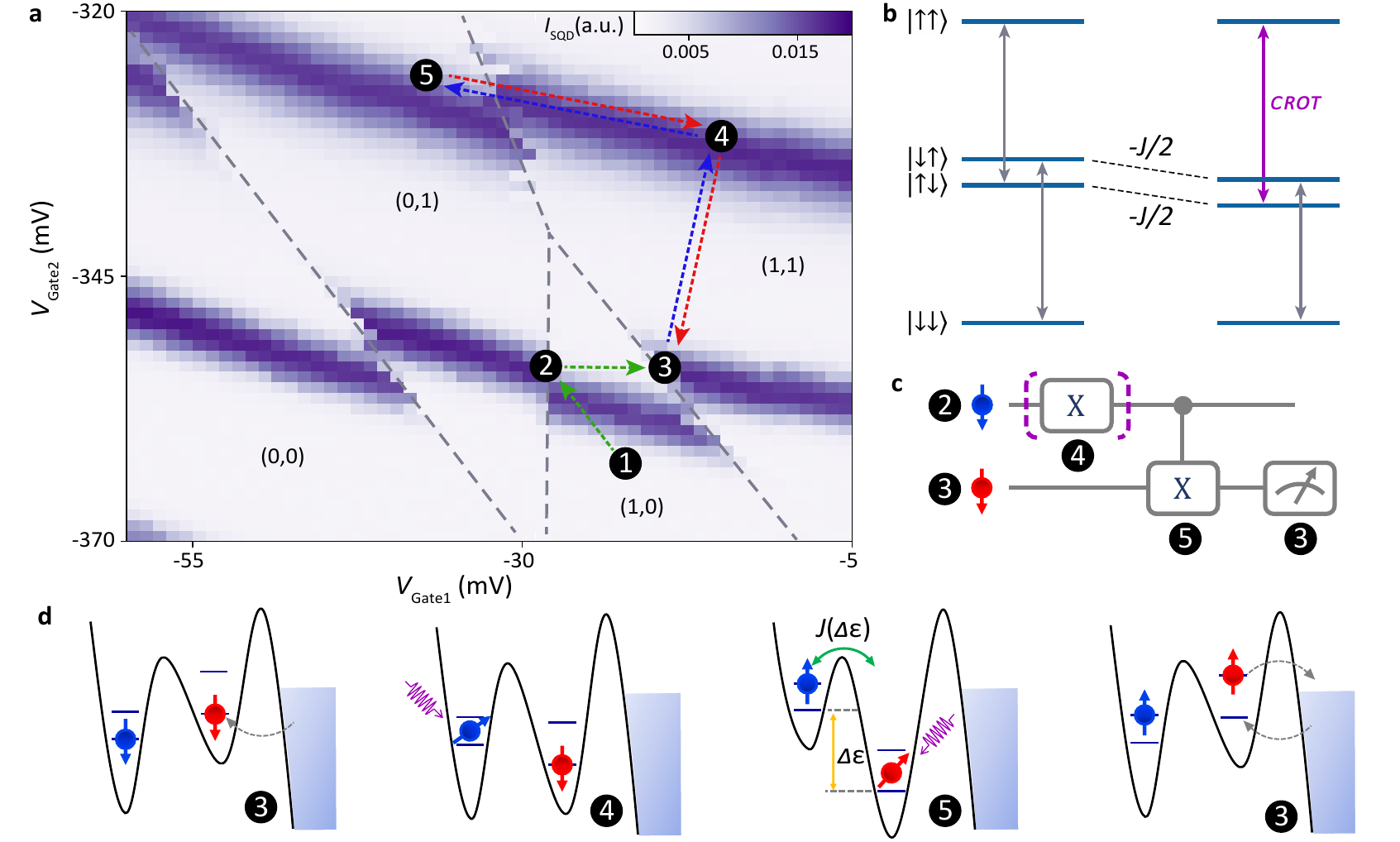}}
\caption{a. Charge stability diagram of the DQD and of the pulsing scheme used in the experiment. The current through a sensing quantum dot is shown in color scale as a function of two gate voltages that control the electrochemical potentials of the two dots. The gate voltages for steps 1 through 5 in the experiment (discussed in the main text) are indicated with black circles, which also appear in panels c and d. We note that the DQD remains in the (1,1) state during step 5 (the CROT gate) because the operation is much faster than the tunneling-out time of Q1. 
b. Spin states of the two-electron system with and without exchange coupling. The spin state of Q1 is mapped onto the spin state of Q2 via a CROT gate based on frequency-selective spin transitions.
c. Quantum circuit for the QND readout procedure. Q1 is used as a logical qubit and as control qubit of the CROT gate, while Q2 is used as ancilla qubit and as target qubit of the CROT gate. An optional $X_{\pi}$ pulse is used to prepare Q1 into the spin-up state following initialization to spin-down. 
d. Schematic representation of the DQD system during the QND measurement protocol. Initialization and readout are implemented by aligning the electrochemical potential of the last electron in dot 2 close to the Fermi energy of the reservoir. The exchange coupling is switched on by detuning the electrochemical potentials of dots 1 and 2. Both electrochemical potentials are varied by applying voltage pulses on the depletion gates that define the DQD confining potential. Spin-flips in single- and two-qubit gates are implemented by EDSR.
}
\label{fig:device}
\end{figure*}

Here we use two electron spin qubits in a double quantum dot (DQD) confined in a Si/SiGe heterostructure. The sample and qubit control techniques are described in detail in Ref.~\cite{watson2018programmable}. In brief, single-qubit gates are realized by electric-dipole spin resonance (EDSR) enabled by the magnetic field gradient from a nearby micromagnet~\cite{pioro2008electrically, kawakami2014electrical}. The micromagnet gradient also causes the resonance frequencies of the two qubits to be well separated. A two-qubit gate is realized by changing the detuning between the chemical potentials of the two dots, which modifies the strength of the exchange interaction, $J$, between the two spins. Due to the interplay of the exchange and the energy difference between the qubits, the energies of the $\ket{01}$ and $\ket{10}$ states are shifted down by $J/2$, because of their coupling to the doubly-occupied singlet state, S(0,2) (Fig.~\ref{fig:device}(b)). Consequently, the EDSR resonance frequency of qubit 2 (Q2) depends on the state of qubit 1 (Q1). By applying an EDSR pulse at the resonance frequency of Q2 corresponding to a particular Q1 state, we obtain a controlled-rotation (CROT) gate:
\begin{align}
\begin{split}
U_{CROT}(\alpha\ket{0}_{Q1} + \beta\ket{1}_{Q1})\ket{0}_{Q2}\\
= \alpha\ket{0}_{Q1}\ket{0}_{Q2} + \beta\ket{1}_{Q1}\ket{1}_{Q2}
\end{split}
\end{align}
Due to the low concentration of nuclear spins in silicon, the mapping from logical qubit to ancilla is close to optimal in every single repetition.
In the QND readout experiment, we choose Q1 (shown in blue in the figures) as the logical qubit and Q2 (shown in red) as the ancilla qubit. We do this for two reasons. First, Q1 has a much longer $T_1$ than Q2 ($T_1 > 50\,\textrm{ms}$ versus $T_1 \approx 1\,\textrm{ms}$), allowing multiple readout cycles of Q1 before significant relaxation occurs. Second, Q2 is physically closer to the Fermi reservoir, which makes it easier to perform destructive readout and reinitialization. Here, the readout of Q2 is performed by detecting spin-selective tunneling to the reservoir with the help of a charge sensor~\cite{elzerman2004single}. The signature of the spin-up state is the appearance of a step in the charge sensor response as only a spin-up electron tunnels out of the quantum dot. A commonly used~\cite{elzerman2004single} and near-optimal~\cite{d'anjou2014optimal} strategy to detect such a step is to compare the peak value $I_p$ of the charge sensor signal during the readout time to a fixed threshold (see Fig.~\ref{fig:soft decoding}c inset for an example of a charge sensor trace). In the data shown below, we infer the spin states using a readout time that minimizes the average single-repetition readout error rate~\cite{supp}.

\begin{figure}[t] 
\center{\includegraphics[width=1.0\linewidth]{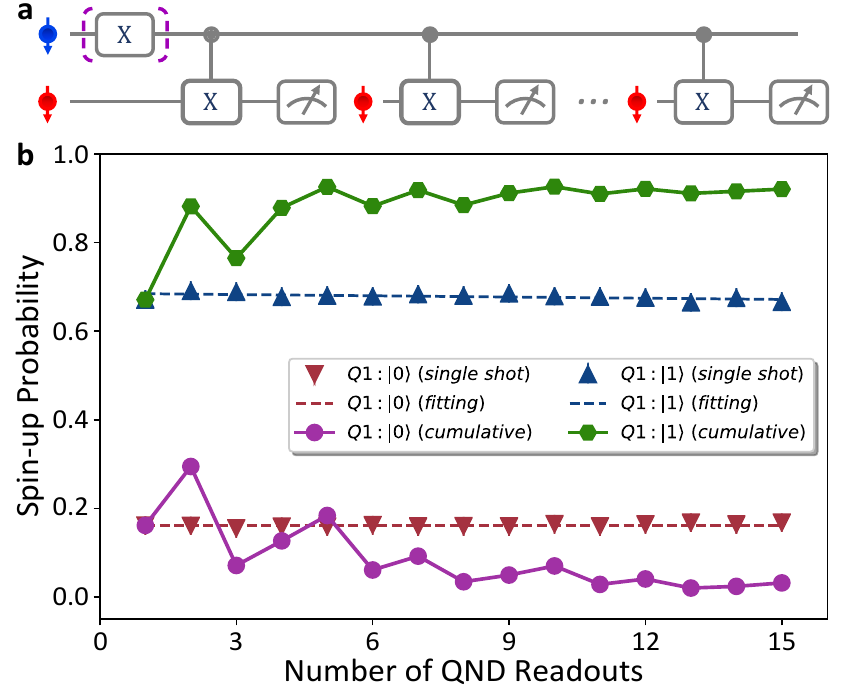}}

\caption{a. Circuit diagram for the repetitive QND readout scheme, using the same labels as in Fig.~\ref{fig:device}c. b. Spin-up probability obtained from individual QND readout cycles (triangles) and from a cumulative weighted majority vote (circles) for preparation of Q1 in state $\ket{1}$ (blue) and in state $\ket{0}$ (red). For the individual readout cycles, the visibility does not improve and in fact slightly decreases due to the finite relaxation time $T_1$ of Q1. By fitting the measured $P_1$ for preparation of Q1 in state $\ket{1}$ to an exponential, we estimate $T_1 = 1.8 \pm 0.6$ s. The cumulative weighted majority vote improves the logical readout visibility as more QND readout cycles are performed.
}
\label{fig:hard decoding}
\end{figure}

We test the QND readout through a protocol whereby voltage pulses applied to two of the quantum dot gate electrodes take the system through the following steps (see Figs.~\ref{fig:device}a,c,d). 1) Empty dot 2, 2) initialize Q1 to the spin-down state via spin relaxation at a hotspot~\cite{srinivasa2013simultaneous}, 3) initialize Q2 in the spin-down state using spin-selective tunneling, 4) an optional single-qubit $\pi$-pulse for initialization of Q1 in the spin-up state, 5) a CROT gate to map the state of Q1 onto the state of Q2 and 6) single-shot readout
of Q2. Step 6 occurs at the same gate voltages as step 3), so at the end of the sequence, Q2 is automatically reinitialized through spin-selective tunneling. In successive QND measurements, steps 1, 2 are omitted, and the optional rotation at step 4 is omitted as well (Fig.~\ref{fig:hard decoding}(a)). Each readout cycle lasts $3.263\,\textrm{ms}$. Because the CROT gate does not affect the state of Q1, successive cycles each yield information on the state of Q1 before the first cycle, as long as Q1 has not been flipped due to relaxation or excitation. Therefore, the readout fidelity of the logical qubit Q1 can be significantly enhanced by repeating the readout cycle. In order to obtain directly the visibility from experiment, we prepare Q1 either in state $\ket{0}$ or in state $\ket{1}$. We then perform up to 15 sequential QND measurements.

\begin{figure*}[t]
\center{\includegraphics[width=1.0\linewidth]{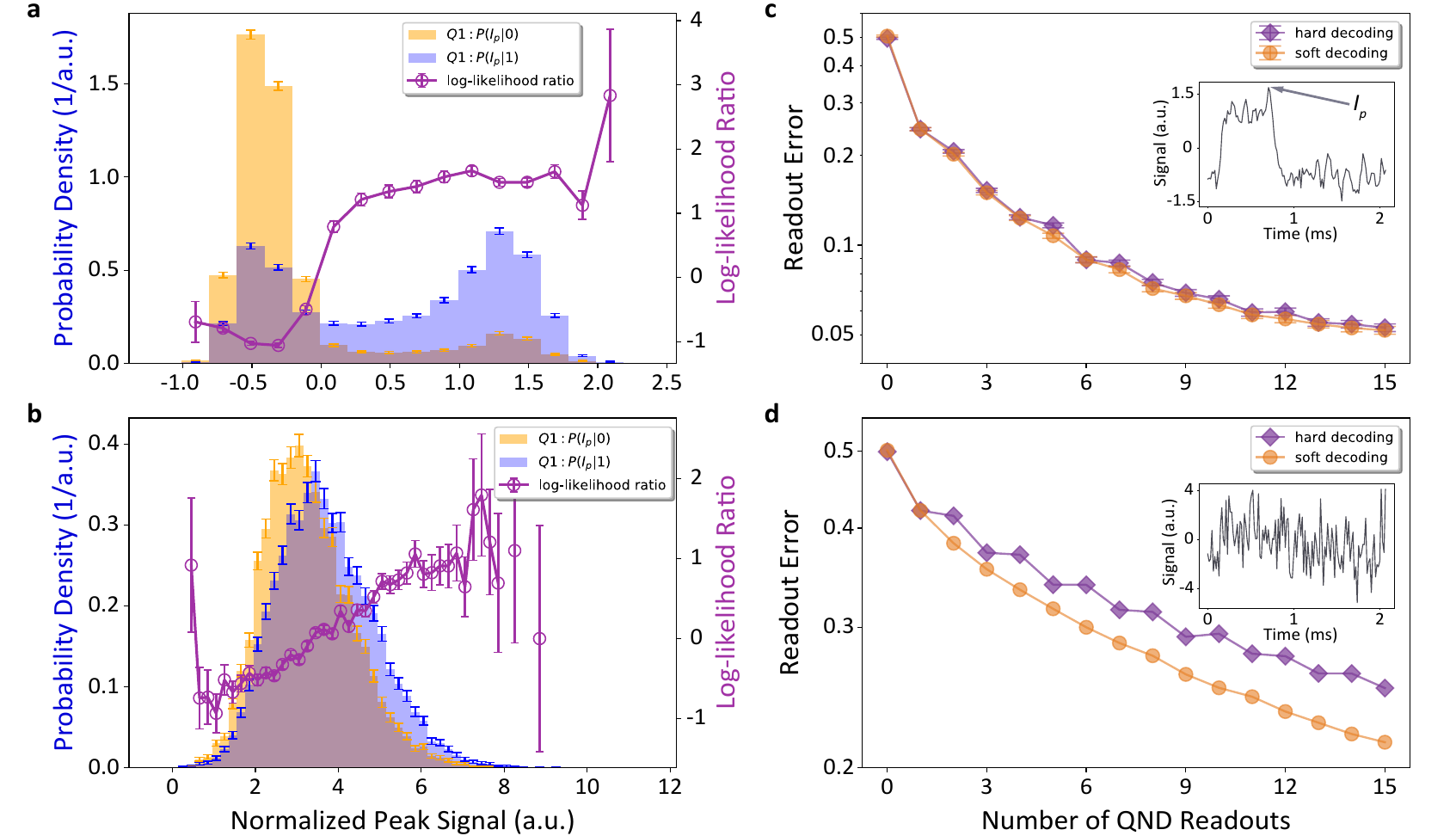}}
\caption{a. Empirically measured distributions $P(I_p|1)$ and $P(I_p|0)$ of the peak sensing dot signal $I_p$ for preparation of Q1 in $\ket{1}$ (blue histograms) and in $\ket{0}$ (yellow histograms), respectively. The distributions are obtained for the readout time that minimizes the single-repetition readout error (see Methods~\cite{supp}). Due to the bimodal features of the two distributions, the log-likelihood ratio $\lambda(I_p)$ is approximately a step function (magenta curve). b. Same dataset as in panel a but with artificially added Gaussian white noise. Here, the two distributions largely overlap. As a result, the log-likelihood ratio $\lambda(I_p)$ is not a step function and different values of $I_p$ on one side of the threshold acquire different weights. This means that if the values of $I_p$ are thresholded at each cycle, useful information is discarded. c, d. Logical readout error rate without and with artificially added white noise. When the log-likelihood ratio $\lambda(I_p)$ approximates a step function, there is little advantage in using soft decoding. When the log-likelihood ratio $\lambda(I_p)$ strongly deviates from a step function, soft-decoding reduces the number of repetitions required to achieve a given error rate. Insets: example readout traces containing a tunnel event without/with added Gaussian white noise.
}
\label{fig:soft decoding}
\end{figure*}

The simplest way to infer the state of Q1 from the repeated readout of Q2 is to perform a majority vote on the readout outcomes~\cite{engel2004measurement}. Ideally, this leads to an exponential suppression of the logical readout error probability $\epsilon^{\textrm{log}}$ with the number of cycles $N$, $\epsilon^{\textrm{log}} \propto \epsilon^N$. Here, $\epsilon$ is the single-repetition readout error rate. 

A slightly more sophisticated approach to inferring the state of Q1 is through a weighted majority vote that accounts for spin relaxation (see Methods for full details~\cite{supp}). Intuitively, the later measurement cycles have accumulated more errors from relaxation than the earlier ones, and are thus given less weight. Fig.~\ref{fig:hard decoding}(b) shows the estimated spin-up probability of Q1 for each individual QND readout cycle as well as for a cumulative weighted majority vote. The cumulative visibility increases from $51.0\pm 0.6\%$ after 1 cycle to $88.9\pm 0.3\%$ after 15 cycles, which corresponds to a cumulative logical fidelity of $94.5\pm 0.2\%$. Here, the logical fidelity is defined as $F^{\textrm{log}} = 1 - \epsilon^{\textrm{log}}$~\cite{supp}. The oscillation in the cumulative spin-up probabilities is due to an even-odd effect from (weighted) majority voting. Due to spin relaxation of Q1, there is a slow decay of the (single-repetition) spin-up probability when Q1 is prepared in $\ket{1}$. Previous $T_1$ measurements on the same device shows that there was no observable decay of Q1 up to 50ms~\cite{watson2018programmable}, consistent with our observations. From the data, it is clear that even higher cumulative fidelities can be achieved using more repetitions (see~\cite{supp} for a detailed discussion).

Further improvements in the readout fidelity for repeated QND readout are possible when taking into account additional information contained in the individual read-out traces. This approach is based on the log-likelihood ratio for the logical state, $\lambda^{\textrm{log}} = \sum_{i=1}^N \lambda(\mathcal{O}_i)$~\cite{kay1998fundamentals} (in the data analysis below, we use a slightly more sophisticated variant, accounting for relaxation of Q1 during the repeated measurements~\cite{supp}). Here, $\mathcal{O}_i$ is the measurement outcome for the $i^{\textrm{th}}$ repetition and $\lambda(\mathcal{O}_i) = \textrm{ln}\left[P(\mathcal{O}_i|1)/P(\mathcal{O}_i|0)\right]$ is the log-likelihood ratio for that outcome. $P(\mathcal{O}_i|1)$ ($P(\mathcal{O}_i|0)$) is the probability to obtain $\mathcal{O}_i$ when the qubit is prepared in $\ket{1}$ ($\ket{0}$).  If $\lambda^{\textrm{log}} > 0$ ($\lambda^{\textrm{log}} < 0$), it is decided that the most likely pre-measurement logical state is $\ket{1}$ ($\ket{0}$). When thresholding, the observable $\mathcal{O}_i$ is either a 1 or a 0, an approach we call `hard decoding'. When considering only a single readout instance, thresholding is optimal. For the repetitive QND readout discussed here, but also for quantum error correction in general, thresholding each individual qubit readout leads to an irreversible loss of information because it discards the level of confidence $\lambda(\mathcal{O}_i)$ that can be assigned to individual readout outcomes~\cite{chase1972class,d'anjou2014soft, dinani2019bayesian,liu2019repetitive}. A better approach is to take an analog variable as the observable $\mathcal{O}_i$ to calculate $\lambda^{\textrm{log}}$. For the readout scheme used here, the peak signal is a good choice~\cite{d'anjou2014optimal}. We refer to this procedure as `soft decoding'. 

The logical readout error probabilities resulting from hard and soft decoding applied to our raw data are plotted in Fig.~\ref{fig:soft decoding}(c) as a function of the number of QND readout cycles. Interestingly, in this instance, the improvement of soft decoding over hard decoding is almost non-existent. This can be understood by examining the empirically measured peak-signal distributions $P(I_p|1)$ and $P(I_p|0)$ at the optimal readout time ($\sim 623\,\mu\textrm{s}$, see Methods~\cite{supp}) shown in Fig.~\ref{fig:soft decoding}(a) along with the log-likelihood ratio $\lambda(I_p)$. Since readout errors caused by noise in the readout traces are small compared to the bit flip errors arising from imperfect CROT operations, ancilla preparation, and spin-to-charge conversion in the readout of the ancilla spin, the peak-signal distributions have clear bimodal features (this is discussed in more detail below). As a result, $\lambda(I_p)$ approximates a step function. This means that all values of $I_p$ on one side of the step are assigned the same level of confidence. It follows that thresholding the values of $I_p$ does not discard much information on the level of confidence in each readout outcome.

For soft decoding to yield an advantage, we must therefore consider situations where $\lambda(I_p)$ is not a step function. One such generic situation occurs in the limit of low SNR for the single-repetition readout. To demonstrate that soft decoding yields an advantage for low SNR, we artificially add Gaussian white noise on top of the experimental readout traces. Fig.~\ref{fig:soft decoding}(d) shows the resulting logical error probabilities for both hard and soft decoding. We see that soft decoding achieves the same logical error rate with 10 repetitions instead of 15 repetitions, a reduction by a third in the number of repetitions. Consequently, a significant amount of readout time may be saved. The reason for this advantage is apparent in Fig.~\ref{fig:soft decoding}(b), which shows the probability distributions $P(I_p|1)$ and $P(I_p|0)$ and the corresponding $\lambda(I_p)$ after adding the noise (optimal readout time $\sim 475\mu\textrm{s}$). Here, the distributions are close to unimodal Gaussians and strongly overlap. This results in a $\lambda(I_p)$ that varies smoothly with $I_p$ such that values of $I_p$ on a given side of the threshold are given different levels of confidence. This is the additional information that yields the soft decoding advantage.

It is important to note that the low-SNR readout is not merely of theoretical interest but is also of great practical relevance. One reason is that it might be difficult to achieve high-SNR in dense qubit arrays where charge sensors, electron reservoirs, or on-chip resonators~\cite{urdampilleta2019gate, zheng2019rapid} are not available, and where only gate-based readout~\cite{pakkiam2018single, west2019gate, urdampilleta2019gate, zheng2019rapid} with SNR limited by a large parasitic capacitance~\cite{pakkiam2018single, west2019gate} is possible. It also opens the possibilities to achieve high-fidelity readout at temperature above 1K~\cite{yang2019silicon, petit2019universal}, which is essential to the integration of quantum dots with cryo-electronics~\cite{charbon2016cryo}. Such situations are precisely the ones where repetitive QND readout may become necessary to achieve low logical readout error rates. Moreover, there are situations where relaxing the constraints on SNR could be beneficial. For instance, it may allow the repetitive readout to operate at lower detector-qubit coupling without loss of fidelity, which could reduce unwanted interference of the detector on the system. Note that these advantages are not limited to the special case of the repetitive QND readout considered here. For instance, quantum error-correcting codes infer error syndromes by using both spatial and temporal redundancy from repeated measurement of multiple ancillas~\cite{kelly2015state}. Both our results and recent theoretical work on continuous-variable quantum error correction~\cite{fukui2017analog, vuillot2019quantum, noh2019encoding, noh2019fault} suggest that soft-decoding of quantum codes could help reduce the number of physical qubits and the number of measurements required to achieve a desired logical error rate.

It must also be emphasized that a low SNR is in general not necessary to benefit from soft decoding. Soft decoding helps when errors arising from noise, as small as they may be, are larger than bit flip errors. This ensures that the log-likelihood ratio changes smoothly instead of step-wise~\cite{d'anjou2014soft}. We discuss ways to engineer such conditions for readout of spin qubits in quantum dots in the supplementary information.

In conclusion, we have performed high-fidelity QND readout of a spin qubit in silicon. The readout fidelity is enhanced by repeatedly mapping the qubit state to a nearby ancilla using a two-qubit gate and measuring the ancilla, from $75.5 \pm 0.3\%$ for a single repetition to $94.5\pm 0.2\%$ for 15 repetitions, and with room for further improvements from additional repetitions~\cite{supp}. We compared two different decoding methods, hard decoding and soft decoding, and discussed the conditions under which soft decoding yields a significant advantage. In the present experiment, hard decoding and soft decoding perform equally well since errors from noise in the readout traces are far less frequent than errors from bit flips. However, with the same rate of bit-flip errors, soft decoding is expected to significantly reduce the number of ancilla measurements required for high-fidelity readout when the SNR is low, as can be the case for gate-based readout in dense qubit arrays, for readout at elevated temperatures, or when SNR must be traded for readout speed.

\textit{Note added.}\textemdash During the preparation of this manuscript, we became aware of related work on repeated QND readout of a silicon spin qubit using the thresholding technique \cite{yoneda2019repetitive}.

Data supporting the findings of this study are available online~\cite{data}
\\

\section*{ACKNOWLEDGMENTS}
This research was sponsored by the Army Research Office (ARO) under grant numbers W911NF-17-1-0274 and W911NF-12-1-0607. The views and conclusions contained in this document are those of the authors and should not be interpreted as representing the official policies, either expressed or implied, of the ARO or the US Government. The US Government is authorized to reproduce and distribute reprints for government purposes notwithstanding any copyright notation herein. Development and maintenance of the growth facilities used for fabricating samples is supported by DOE (DE- FG02-03ER46028). B.D. acknowledges funding from NSERC. W.A.C. acknowledges funding from NSERC, CIFAR, and FRQNT. We acknowledge the use of facilities supported by NSF through the University of Wisconsin-Madison MRSEC (DMR-1121288). We acknowledge useful discussions with Matthew G. House, and the members of the Vandersypen group, and technical assistance by R. Schouten and R. Vermeulen.

\bibliography{main}



\end{document}


\beginsupplement

\title{Supplemental material: Repetitive quantum non-demolition measurement and soft decoding of a silicon spin qubit}

\author{Xiao~Xue}
 \thanks{These authors contributed equally to this work}
\affiliation{QuTech and Kavli Institute of Nanosicence, Delft University of Technology, Lorentzweg 1, 2628 CJ Delft, The Netherlands}
\author{Benjamin~D'Anjou}
 \thanks{These authors contributed equally to this work}
\affiliation{Department of Physics, University of Konstanz, D-78457 Konstanz, Germany}
\author{Thomas~F.~Watson}
\affiliation{QuTech and Kavli Institute of Nanosicence, Delft University of Technology, Lorentzweg 1, 2628 CJ Delft, The Netherlands}
\author{Daniel~R.~Ward}
\affiliation{University of Wisconsin-Madison, Madison, WI 53706, USA}
\author{Donald~E.~Savage}
\affiliation{University of Wisconsin-Madison, Madison, WI 53706, USA}
\author{Max~G.~Lagally}
\affiliation{University of Wisconsin-Madison, Madison, WI 53706, USA}
\author{Mark~Friesen}
\affiliation{University of Wisconsin-Madison, Madison, WI 53706, USA}
\author{Susan~N.~Coppersmith}
 \thanks{Present address: School of Physics, University of New South Wales, Sydney NSW 2052, Australia.}
\affiliation{University of Wisconsin-Madison, Madison, WI 53706, USA}
\author{Mark~A.~Eriksson}
\affiliation{University of Wisconsin-Madison, Madison, WI 53706, USA}
\author{William~A.~Coish}
\affiliation{Department of Physics, McGill University, Montreal, Quebec H3A 2T8, Canada}
\author{Lieven~M.~K.~Vandersypen}
 \email{Correspondence author: l.m.k.vandersypen@tudelft.nl}
\affiliation{QuTech and Kavli Institute of Nanosicence, Delft University of Technology, Lorentzweg 1, 2628 CJ Delft, The Netherlands}

\date{\today}

\maketitle

\section{Bayesian inference \label{sec:bayesianInference}}

In this section, we give an algorithm to efficiently perform optimal Bayesian inference of the logical qubit state using the empirically determined statistics of the repetitive readout.

\subsection{Repetitive readout}

Suppose that the logical qubit is repetitively read out $N$ times, with each repetition having a duration $\delta t_{\textrm{rep}}$. We may consider the state of the logical qubit at the discrete times:
\begin{align}
	t_k = k\,\delta t_{\textrm{rep}}, \;\;\; k=0,1,...,{N},
\end{align}
with the $k^{\textrm{th}}$ repetition taking place in the interval $\left[t_{k-1},t_k\right]$. Here, the coherence of the logical qubit in the computational basis plays no role in the statistics of the measurement. For the present purposes, we may therefore model the time evolution of the logical qubit classically. The classical state of the logical qubit at time $t_k$ is labeled $\staX_k$. The state evolution of the logical qubit up to $k \leq N$ repetitions follows the stochastic time series:
\begin{align}
	\StaX_{k} = \left\{ \staX_0, \staX_1, \dots, \staX_k \right\}.
\end{align}
In reality, each individual readout is noisy. Thus, the state $\staX_k$ at each repetition is not directly recorded. Instead, a noisy observation $\obs_k$ is recorded. This gives the observation sequence:
\begin{align}
	\Obs_k = \left\{\obs_0, \obs_1, \dots ,\obs_{k-1}\right\}.
\end{align}
Note that in general, the observations $\obs_k$ need not be scalar.

\subsection{Maximum-likelihood decision}

Our goal is to infer the most likely initial state of the logical qubit from the sequence $\Obs_k$ of noisy repetitive readout outcomes. This is most easily done by calculating the posterior probability ratio:
\begin{align}
	\frac{P(\staX_0 = 1|\Obs_k)}{P(\staX_0 = 0|\Obs_k)} = \frac{P(\Obs_k|\staX_0 = 1)}{P(\Obs_k|\staX_0 = 0)} \frac{P(\staX_0 = 1)}{P(\staX_0 = 0)}.
\end{align}
The initial state is most likely $1$ if the ratio is larger than unity, and it is most likely $0$ if the ratio is smaller than unity. In the absence of prior information on the logical qubit state, $P(\staX_0 = 0) = P(\staX_0 = 1) = 1/2$, the above {\it maximum a posteriori} decision~\cite{kay1998} reduces to calculating the log-likelihood ratio:
\begin{align}
	\lambda^{\textrm{log}}_k = \ln \frac{P(\Obs_k|\staX_0 = 1)}{P(\Obs_k|\staX_0 = 0)}. \label{eq:logicalLogLikelihoodRatio}
\end{align}
The initial state is now most likely $1$ if $\lambda^{\textrm{log}}_k > 0$, and it is most likely $0$ if $\lambda^{\textrm{log}}_k < 0$. This results in a maximum-likelihood decision~\cite{kay1998}.

\subsection{Logical readout error rate \label{eq:logicalErrorRate}}

The average logical readout error rate $\epsilon$ is given by
\begin{align}
	\epsilon^{\textrm{log}} = \frac{1}{2}\left(\epsilon^{\textrm{log}}_1 + \epsilon^{\textrm{log}}_0\right).
\end{align}
Here, $\epsilon^{\textrm{log}}_1$ and $\epsilon^{\textrm{log}}_0$ are the error rates conditioned on preparation of the logical qubit in states $1$ and $0$ at time $t_0$, respectively. These are given by:
\begin{align}
\begin{split}
	\epsilon^{\textrm{log}}_1 = P(\lambda^{\textrm{log}}_k < 0| \staX_0 = 1), \\
	\epsilon^{\textrm{log}}_0 = P(\lambda^{\textrm{log}}_k > 0| \staX_0 = 0).
\end{split}	
\end{align}
An experimental estimate of the logical readout error rate $\epsilon^{\textrm{log}}_1$ ($\epsilon^{\textrm{log}}_0$) is obtained by preparing the logical qubit in state $\ket{1}$ ($\ket{0}$) $10^4$ times, measuring the record $\Obs_k$ and calculating $\lambda^{\textrm{log}}_k$ for each attempt, and counting the number of times where $\lambda_k^{\textrm{log}} < 0$ ($\lambda_k^{\textrm{log}} > 0$). Finally, we note that the average readout fidelity quoted in the main text is defined as $F^{\textrm{log}} = 1 - \epsilon^{\textrm{log}}$, while the logical visibility is $\mathcal{V}^{\textrm{log}} = 1 - 2\epsilon^{\textrm{log}}$.

\subsection{Calculating the log-likelihood ratio}

To calculate $\lambda^{\textrm{log}}_k$, the statistics of the observations $\Obs_k$ given the initial state of the logical qubit must be known. More precisely, the probability distributions $P(\Obs_k|\staX_0 = 1)$ and $P(\Obs_k|\staX_0 = 0)$ appearing in Eq.~\eqref{eq:logicalLogLikelihoodRatio} must be calculated. In the following, we show how to model and calculate these distributions for the repetitive readout using the theory of hidden Markov models~\cite{hann2018,nakajima2019,elder2019,bultink2019,yoneda2019}. The hidden Markov model provides a direct connection between the single-repetition probability distributions, $P(\obs_k|\staX_k)$, and the probability distributions for the full measurement record, $P(\Obs_k|\staX_0)$, accounting for the dynamics of the logical qubit.  From $P(\Obs_k|\staX_0)$, the log-likelihood ratio can be evaluated directly using Eq.~\eqref{eq:logicalLogLikelihoodRatio}.  As described in Sec.~\ref{sec:repetitiveStatistics}, following Eq.~\eqref{eq:softNoiseVector}, here we have determined $P(\obs_k|\staX_k)$ empirically. However, note that these distributions can also be found from an appropriate dynamical model of the readout. See for instance Refs.~\cite{gambetta2007,gammelmark2014,ng2014,danjou2014,danjou2016} and others, where hidden Markov models are used to determine $P(\obs_k|\staX_k)$ at the level of a single repetition.

\subsubsection{Hidden Markov models}

As discussed in the main text, the logical qubit state evolves during the measurement via spin relaxation on a time scale $T_1$. Such a process is Markovian in the sense that the statistics of the state at time $t_{k+1}$ are fully determined by the state at time $t_{k}$:
\begin{align}
	P(\staX_{k+1}|\StaX_k) = P(\staX_{k+1}|\staX_k). \label{eq:markovCondition}
\end{align}
Because the ancilla qubit is reinitialized after each repetition, the noisy observations $\obs_k$ are independent from each other and depend only on the state of the logical qubit at time $t_k$:
\begin{align}
	P(\Obs_k|\StaX_k) =  \prod_{l=0}^{k-1} P(\obs_l|\staX_l). \label{eq:whiteNoise}
\end{align}
In other words, the observation noise is white. Finally, prior knowledge about the system state at each time is specified by the prior probability distribution $P(\staX_k)$ for the state $\staX_k$ at each time $t_k$. The set $\left\{P(\staX_{k+1}|\staX_k), P(\obs_k|\staX_k), P(\staX_k)\right\}$ defines a `Hidden Markov Model'.

\subsubsection{Forward filtering}

For the hidden Markov models discussed above, the log-likelihood ratio may be calculated with the help of an iterative forward filtering algorithm for the logical qubit state~\cite{gambetta2007,gammelmark2013,gammelmark2014,danjou2014,danjou2016,hann2018,nakajima2019,elder2019,bultink2019,yoneda2019}. Forward filtering consists in calculating the probability distribution of the logical qubit state $\staX_k$ at time $t_k$ given all previous observations. We denote this probability distribution as
\begin{align}
	\varrho_k(\staX_k) = P(\staX_k|\Obs_{k}).
\end{align}
Using Bayes' rule, the distribution $\varrho_k(\staX_k)$ may be rewritten as
\begin{align}
	\varrho_k(\staX_k) = \frac{P(\staX_k,\Obs_{k})}{P(\Obs_{k})} = \frac{P(\staX_k,\Obs_{k})}{\sum_\staY P(\staY,\Obs_{k})} = \frac{\ell_k(\staX_k)}{\sum_\staY \ell_k(\staY)}. \label{eq:likelihoodToState}
\end{align}
Here,
\begin{align}
	\ell_k(\staX_k) = P(\staX_k,\Obs_k)
\end{align}
is the joint probability distribution of the state $\staX_k$ and of the previous observations. One advantage of calculating $\ell_k(\staX_k)$ instead of $\varrho_k(\staX_k)$ is that $\ell_k(\staX_k)$ obeys a linear recurrence relation while $\varrho_k(\staX_k)$ obeys a non-linear recurrence relation (see Sec.~\ref{sec:filteringEquations}).

We note that the denominator in Eq.~\eqref{eq:likelihoodToState} is the likelihood function
\begin{align}
	\mathcal{L}_k = P(\Obs_k) = \sum_{\staX_k} \ell_k(\staX_k). \label{eq:likelihoodFunction}
\end{align}
Eq.~\eqref{eq:likelihoodToState} now takes the form
\begin{align}
	\ell_k(\staX_k) = \varrho_k(\staX_k) \times \mathcal{L}_k. \label{eq:likelihoodToState2}
\end{align}
It is convenient to introduce the vector notation
\begin{align}
\begin{split}
	\ketR{\varrho_k} = \sum_\staX \varrho_k(\staX) \ketR{\staX}, \;\;\;\ketR{\ell_k} = \sum_\staX \ell_k(\staX) \ketR{\staX},
\end{split} 
\end{align}
where $\left\{\ketR{\staX}\right\}$ is a set of basis vectors representing the classical logical qubit states $x$. Eq.~\eqref{eq:likelihoodToState2} is then rewritten as
\begin{align}
	\ketR{\ell_k} = \mathcal{L}_k  \times \ketR{\varrho_k}.\label{eq:likelihoodToStateVec}
\end{align}
The likelihood function, Eq.~\eqref{eq:likelihoodFunction}, may compactly be written as
\begin{align}
	\mathcal{L}_k = \Tr \ketR{\ell_k}. \label{eq:likelihoodFunctionVec}
\end{align}
Here, the trace $\Tr$ of a vector is the sum of its elements. Choosing the basis $\left\{\ketR{1},\ketR{0}\right\}$, the likelihood function may be calculated for the initial states $\staX_0 = 1$ and $\staX_0 = 0$ of the logical qubit by setting $\ketR{\varrho_0} = \ketR{\ell_0} = (1,0)^\textrm{T}$ and $\ketR{\varrho_0} = \ketR{\ell_0} = (0,1)^\textrm{T}$, respectively. This enables the maximum-likelihood decision discussed following Eq.~\eqref{eq:logicalLogLikelihoodRatio}.

\subsubsection{The filtering equations \label{sec:filteringEquations}}

For completeness, we now derive the recurrence relation for forward filtering of hidden Markov models. We note that
\begin{align}
\begin{split}
	\ell_{k+1}(\staX_{k+1}) = P(\staX_{k+1},\Obs_{k+1}) &= \sum_{\staX_k} P(\staX_{k+1},\Obs_{k+1}|\staX_k)P(\staX_k) \\
	&= \sum_{\staX_k} P(\Obs_{k+1}|\staX_{k+1},\staX_k)P(\staX_{k+1}|\staX_k)P(\staX_k).
\end{split}
\end{align}
Next we recall that 1) $\Obs_{k+1} = \left\{\obs_1,\obs_2,\dots,\obs_k \right\}$ does not depend on $\staX_{k+1}$ by causality and that 2) the observation noise is white, Eq.~\eqref{eq:whiteNoise}. Thus, we have
\begin{align}
\begin{split}
	\ell_{k+1}(\staX_{k+1}) &= \sum_{\staX_k} P(\Obs_{k+1}|\staX_k)P(\staX_{k+1}|\staX_k)P(\staX_k) \\
	&= \sum_{\staX_k} P(\obs_k|\staX_k)P(\Obs_k|\staX_k)P(\staX_{k+1}|\staX_k)P(\staX_k) \\
	&= \sum_{\staX_k} P(\obs_k|\staX_k)P(\staX_{k+1}|\staX_k)\ell_k(\staX_k).
\end{split}
\end{align}
This recurrence relation may be written in vector form
\begin{align}
   \ketR{\ell_{k+1}} = V_k(\obs_k) \ketR{\ell_k}.
\end{align}
Here, we have introduced the matrix $V_k(\obs_k)$ with elements
\begin{align}
   &\braR{x}V_k(\obs_k)\ketR{\staY} = P(\staX_{k+1}|\staY_k) P(\obs_k|\staY_k) = w_k^{\staX \staY}\, \mathcal{P}_k^y(\obs_k),
\end{align}
and we have defined
\begin{align}
\begin{split}
   &w_k^{\staX \staY} = P(\staX_{k+1}|\staY_k), \\
   &\mathcal{P}_k^\staX(\obs_k) = P(\obs_k|\staX_k). \label{eq:modelMatrices}
\end{split}
\end{align}
The matrix $w_k^{\staX \staY}$ describes the transition probabilities for the evolution of the logical qubit state and the vector $\mathcal{P}_k^\staX(\obs_k)$ describes the observation noise for each logical qubit state. Note that there exist corresponding recurrence relations for the state vector $\ketR{\varrho_k}$ and for the likelihood function $\mathcal{L}_k$. They take the form
\begin{align}
	\ketR{\varrho_{k+1}} = \frac{1}{\mathcal{N}_{k+1}} \ketR{\tilde{\varrho}_{k+1}}, \;\;\; \mathcal{L}_{k+1} =  \mathcal{N}_{k+1} \times \mathcal{L}_k,
\end{align}
where
\begin{align}
	\ketR{\tilde{\varrho}_{k+1}} = V_k(\obs_k) \ketR{\varrho_k}, \;\;\; \mathcal{N}_{k+1} = \Tr\ketR{\tilde{\varrho}_{k+1}}.
\end{align}

\subsubsection{Numerical algorithm \label{sec:numericalAlgorithm}}

We now provide an efficient numerical algorithm for forward filtering. The algorithm simultaneously calculates the probability vector $\ketR{\varrho_k}$ and the log-likelihood function $\ln \mathcal{L}_k$ as follows:
\begin{enumerate}
	\item Set $k=0$.
	\item Calculate the matrix $V_k(\obs_k)$.
	\item Calculate $\ketR{\tilde{\varrho}_{k+1}} = V_k(\obs_k) \ketR{\varrho_k}$.
	\item Calculate the norm $\mathcal{N}_{k+1} = \Tr \ketR{\tilde{\varrho}_{k+1}}$.
	\item Update the probability distribution $\ketR{\varrho_{k+1}} = \ketR{\tilde{\varrho}_{k+1}}/\mathcal{N}_{k+1}$.
	\item Update the log-likelihood ratio with $\ln\mathcal{L}_{k+1} = \ln\mathcal{L}_{k} + \ln \mathcal{N}_{k+1}$.
	\item Increase $k$ by one and start again.
\end{enumerate}
Note that we are only interested in estimating the initial state of the logical qubit. Therefore, the matrix $V_k(\obs_k)$ may be normalized at each step by any constant factor independent of the qubit state without affecting the maximum-likelihood estimate. In some cases, this may prevent the values of the log-likelihood function from becoming too large. 

\section{Repetitive readout statistics \label{sec:repetitiveStatistics}}

The hidden Markov model relevant for the repetitive readout discussed in the main text is obtained by specifying the transition matrix $w_k$ and the noise vector $\mathcal{P}_k$ appearing in Eq.~\eqref{eq:modelMatrices}.

As discussed in the main text, the qubit undergoes relaxation to its spin ground state on a time scale $T_1$. We obtain the value of $T_1$ by simultaneously fitting the measured single-repetition probabilities $P_1$ and $P_0$ shown in Fig.~2b to an expression of the form
\begin{align}
\begin{split}
P_1(t) &= A \epsilon^{-t/T_1} + B, 
\end{split}
\end{align}
for the case when Q1 is prepared in $\ket{1}$ before the first measurement cycle, and of the form
\begin{align}
P_0(t) &= B.
\end{align}
for the case when Q1 is initially prepared in $\ket{0}$. Here, $t$ is the total readout time after the beginning of the repetitive readout and $A$ and $B$ are constants. The fit is shown in Fig.~2b and yields $T_1 = 1.8 \pm 0.6 s$. For a relaxation process, the transition matrix $w_k^{\staX \staY}$ in the basis $\left\{\ketR{1},\ketR{0}\right\}$ takes the form:
\begin{align}
	w_k =
	\textrm{exp}
	\left[
	\left(
	\begin{array}{cc}
	-1 & 0\\
	 1 & 0
	\end{array}
	\right)
	\frac{\delta t_{\textrm{rep}}}{T_1}
	\right]. \label{eq:lindbladian}
\end{align}
Here, $\delta t_{\textrm{rep}} = 3.263\,\textrm{ms}$.

The noise vector $\mathcal{P}_k^y(\obs_k)$ describing the distribution of the outcome $\obs_k$ for each logical qubit state is given by
\begin{align}
	\mathcal{P}_k(\obs_k) =
	\left(
	\begin{array}{c}
	P(\obs_k|\staX_k = 1) \\
	P(\obs_k|\staX_k = 0)
	\end{array}
	\right). \label{eq:softNoiseVector}
\end{align}
Here, $P(\obs_k|\staX_k = 1)$ and $P(\obs_k|\staX_k = 0)$ are the probability distributions of the readout outcome $\obs_k$ for preparation of the logical qubit in states $1$ and $0$, respectively. For the soft decoding procedure discussed in the main text, the readout outcome $\obs_k$ is taken to be the peak signal $I_p$ (see Sec.~\ref{sec:singleRepetition}). The distributions of outcomes conditioned on the logical qubit state are then simply the empirically observed distributions of the peak signal $P(I_p|\staX_k = 1)$ and $P(I_p|\staX_k = 0)$ displayed in Figs.~3a-b of the main text. For the hard decoding procedure, the readout outcome $\obs_k$ is taken to be the binary value $1$ or $0$ obtained by thresholding the peak signal at each repetition. The distributions of outcomes conditioned on the logical qubit state are then given by the conditional single-repetition readout error rates $\epsilon_1$ and $\epsilon_0$:
\begin{align}
\begin{array}{ll}
	P(1|\staX_k = 1) = 1 - \epsilon_1,  &P(1|\staX_k = 0) = \epsilon_0, \\
	P(0|\staX_k = 1) = \epsilon_1,  &P(0|\staX_k = 0) = 1-\epsilon_0 . \label{eq:hardNoiseVector}
\end{array}
\end{align}
The procedure used to obtain the distributions $P(I_p|\staX_k = 1)$ and $P(I_p|\staX_k = 0)$ and the conditional single-repetition readout error rates $\epsilon_1$ and $\epsilon_0$ is detailed in Sec.~\ref{sec:singleRepetition}.

\section{Single-repetition readout calibration \label{sec:singleRepetition}}

\begin{figure} 
\center{\includegraphics[width=\columnwidth]{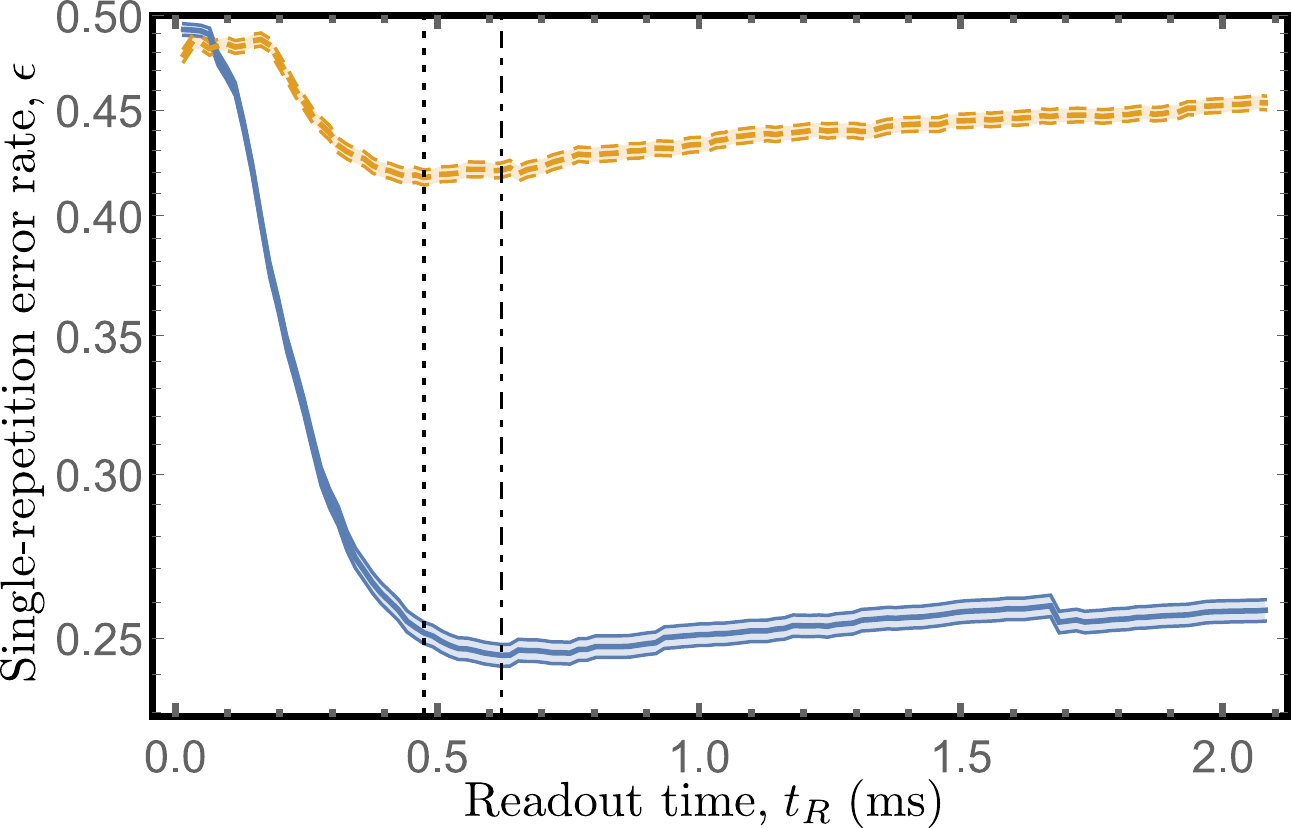}}
\caption{Average single-repetition readout error rate $\epsilon$ as a function of readout time $t_R$ in the absence of added noise (solid blue line) and in the presence of added noise (dashed yellow line). The optimal readout time in the absence (presence) of added noise is indicated by the dot-dashed (dotted) line. In both cases, the shaded areas give the statistical uncertainty (see text). Finally, we note that the small discontinuity in the blue curve at $t_R \approx 1.7\,\textrm{ms}$ is a consequence of the finite histogram bin size and has no particular physical significance.}
\label{fig:figS1}
\end{figure}

Since all repetitions are identical, we use the first repetition to calibrate the single-repetition readout. At $t_0 = 0$, the logical qubit is prepared $10^4$ times in state $\staX_0 = 1$ and $10^4$ times in state $\staX_0 = 0$. For each preparation, a readout trace $I(t)$ is recorded for a total time of $2.01\,\textrm{ms}$ in steps of $\delta t = 16.38\,\mu\textrm{s}$. For all readout times $t_R \leq 2.01\,\textrm{ms}$, we extract the peak signal $I_p = \max_t I(t)$~\cite{elzerman2004} and construct the probability distributions of $I_p$ conditioned on the logical qubit state, $P(I_p|\staX_0=1)$ and $P(I_p|\staX_0=0)$. For a given readout time $t_R$, single-repetition readout is performed by calculating the single-repetition log-likelihood ratio
\begin{align}
	\lambda(I_p) = \ln \frac{P(I_p|\staX_0=1)}{P(I_p|\staX_0=0)}.
\end{align}
If $\lambda(I_p) > 0$ [$\lambda(I_p) < 0$], we decide that the qubit state is most likely $\ket{1}$ [$\ket{0}$]. Therefore, the single-repetition readout error rates $\epsilon_1$ and $\epsilon_0$ for each state are obtained from marginals of the distributions $P(I_p|\staX_0 = 1)$ and $P(I_p|\staX_0 = 0)$ as follows:
\begin{align}
\begin{split}
	\epsilon_1 = P(\lambda^{\textrm{log}}<0|\staX_0 = 1) = \sum_{\left\{I_p | \lambda(I_p)<0\right\}} P(I_p|\staX_0 = 1), \\
	\epsilon_0 = P(\lambda^{\textrm{log}}>0|\staX_0 = 0)= \sum_{\left\{I_p|\lambda(I_p)>0\right\}} P(I_p|\staX_0 = 0), \label{eq:conditionalSingleShot}
\end{split}
\end{align}
and the average single-repetition readout error rate is given by:
\begin{align}
	\epsilon = \frac{1}{2}\left(\epsilon_1 + \epsilon_0\right)
\end{align}
The value of $\epsilon$ is plotted in Fig.~\ref{fig:figS1} as a function of readout time in the absence and in the presence of artificially added noise. We choose the value of $t_R$ that minimizes $\epsilon$. The histograms $P(I_p|\staX_0 = 1)$ and $P(I_p|\staX_0 = 0)$ and the error rates $\epsilon_1$ and $\epsilon_0$ corresponding to that optimal value of $t_R$ are those used throughout the main text. In particular, they are used to perform the Bayesian analysis detailed in Sec.~\ref{sec:repetitiveStatistics}. Without adding artifical noise to the readout traces, we find $\epsilon_1 = 32.9 \pm 0.5\%$ and $\epsilon_0 = 16.2 \pm 0.4\%$ at an optimal readout time $t_R = 623\,\mu\textrm{s}$. After adding noise, we find $\epsilon_1 = 41.1\pm 0.5 \%$ and $\epsilon_0 = 42.3\pm 0.5\%$ at an optimal readout time $t_R = 475\,\mu\textrm{s}$. The error bars on $\epsilon_1$ and $\epsilon_0$ are given by the standard deviation of the corresponding binomial error process with $N = 10^4$ samples, $\delta \epsilon_{1} = \sqrt{\epsilon_{1}(1-\epsilon_{1})/N}$ and $\delta \epsilon_{0} = \sqrt{\epsilon_{0}(1-\epsilon_{0})/N}$.

Note that we have assumed perfect preparation of the logical qubit state throughout. For this device, preparation errors have been reported~\cite{watson2018} to be $\sim 1\%$, smaller than the single-repetition readout errors reported in Sec.~\ref{sec:singleRepetition}. In Sec.~\ref{sec:theoreticalSimulation}, we argue that the average preparation error is in fact around $2-3\%$ ($4-5\%$ for state $\ket{1}$ and $< 1\%$ for state $\ket{0}$) for this run of the experiment. Therefore, the empirically measured single-repetition readout distributions in Figs.~3a-b are close to what they would be without preparation errors and lead to decoding procedures (hard and soft) that are close to optimal. In Sec.~\ref{sec:theoreticalSimulation}, we give rough estimates for the expected logical readout fidelity that would be measured in the absence of preparation errors.

\section{Theoretical simulation \label{sec:theoreticalSimulation}}

To explore the full potential of the repetitive readout discussed in the main text, we perform a numerical simulation. For each qubit state, we randomly sample $10^4$ outcome sequences according to the hidden Markov model described in Sec.~\ref{sec:repetitiveStatistics}. We use the experimentally extracted value of $T_1$ and the empirically measured peak-signal distributions of Fig.~3a as input parameters. A Monte-Carlo estimate of the average logical readout error rate $\epsilon^{\textrm{log}}$ is obtained by thresholding the simulated outcome sequences as described in Sec.~\ref{sec:singleRepetition} and then applying the hard decoding procedure detailed in Sec.~\ref{sec:bayesianInference}.
\begin{figure} 
\center{\includegraphics[width=0.8\columnwidth]{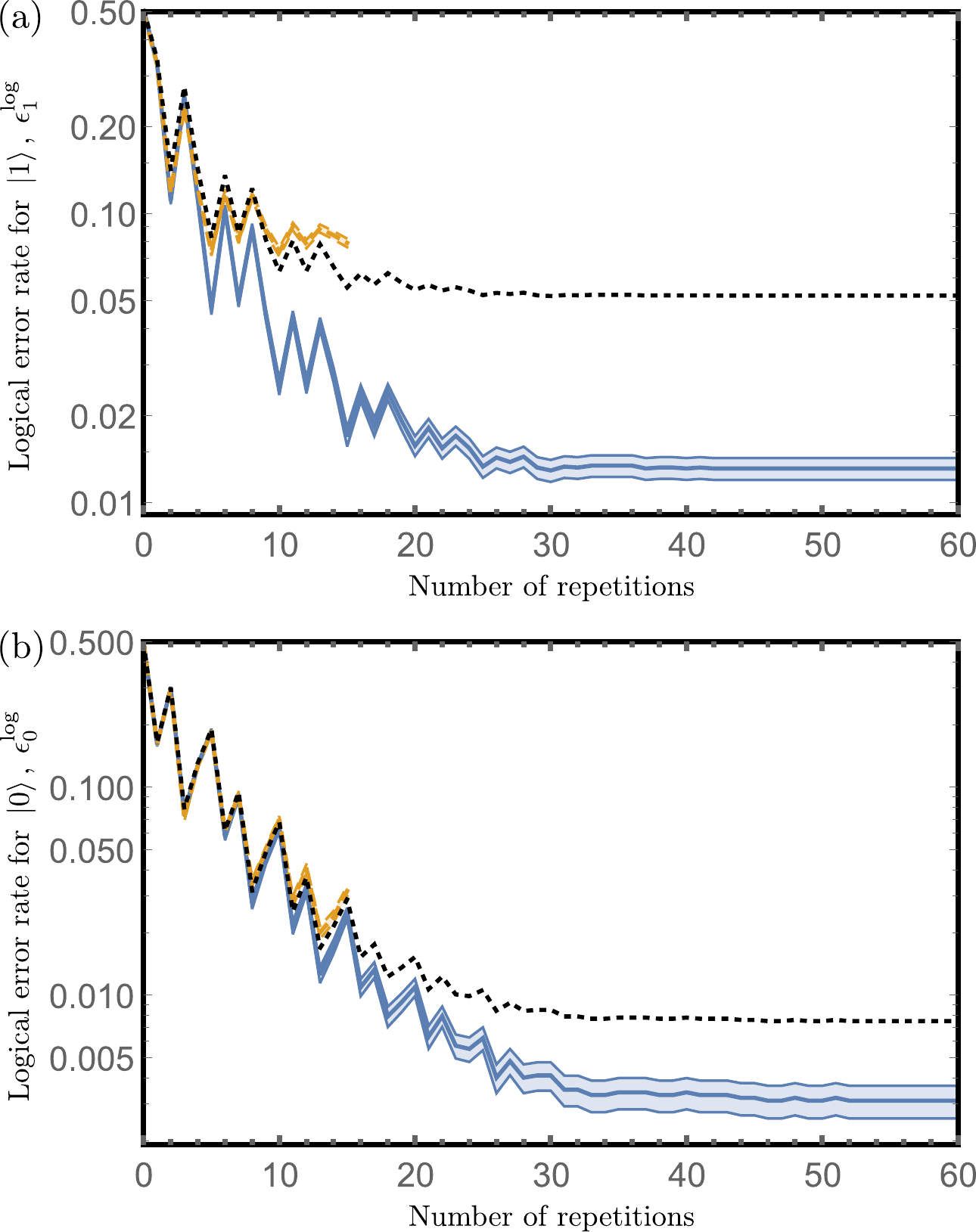}}
\caption{Simulated logical readout error rate (solid blue line) as a function of the number of repetitions for (a) state $\ket{1}$ and (b) state $\ket{0}$. In both cases, the experimentally measured logical readout error rate is shown for comparison (dashed yellow line). The bands give the statistical error associated with the binomial statistics of the sampled errors as explained in Sec.~\ref{sec:singleRepetition}. Moreover, the dotted black curves give the simulated curves corrected to fit the experimental curves accounting for finite preparation errors using Eq.~\eqref{eq:errorComposition}. The best fit values for the preparation errors are $\eta_1 = 4\%$ for state $\ket{1}$ and $\eta_0 = 0.44\%$ for state $\ket{0}$, resulting in an average preparation error of $\eta = 2.2\%$. }
\label{fig:figS2}
\end{figure}

The conditional logical readout error rates $\epsilon_1^{\textrm{log}}$ and $\epsilon_0^{\textrm{log}}$ are plotted in Fig.~\ref{fig:figS2} as a function of the number of repetitions. The experimental results of Fig.~3c are shown for comparison. It is clear from Fig.~\ref{fig:figS2} that the simulated logical readout error rate underestimates the experimentally measured readout error rate as the number of repetitions increases. We believe that this is due to the logical readout error rate becoming comparable to preparation errors for Q1. This assumption is consistent with the simulation and experiment agreeing at low repetition number, where the logical readout error rate is much larger than the expected preparation error (of the order of $1\%$~\cite{watson2018}). In fact, it is possible to estimate the preparation error $\eta$ by fitting the simulated error rate $\epsilon^{\textrm{sim}}$ to the experimentally measured error rate $\epsilon^{\textrm{log}}$ using the error composition relation
\begin{align}
  \epsilon^{\textrm{log}} = (1-2\eta)\epsilon^{\textrm{sim}} + \eta. \label{eq:errorComposition}
\end{align}
We find that the simulation and experiment agree for a preparation error of $\eta_1 \approx 4-5\%$ for state $\ket{1}$ and of $\eta_0 < 1\%$ for state $\ket{0}$, giving an average preparation error of approximately $\eta \approx 2-3\%$. The simulation results suggest that in the absence of preparation errors, the measured fidelity could reach $\sim 98\%$ after 15 repetitions and saturate at $>99\%$ for more than 30 repetitions.

\section{Engineering Gaussian distributions}

It is known that if $P(I_p|1)$ and $P(I_p|0)$ are unimodal Gaussian distributions, soft decoding can reduce the number of readout cycles by up to a factor of two for arbitrarily large readout SNR. For this advantage to exist, the `bit-flip' conversion errors creating the bimodal features of the measured probability distributions must be small enough that the errors are dominated by the Gaussian noise~\cite{danjou2014-2}. Understanding the origin of the conversion errors is thus of great importance. Here, the `bit-flip' conversion errors may arise from an imperfect CROT gate, imperfect ancilla preparation, or imperfect spin-to-charge conversion in the readout of Q2. The CROT and preparation errors will be generically suppressed as the control and  preparation fidelities are improved. The imperfect spin-to-charge conversion in the readout of Q2 may be the most challenging to overcome. In what follows, we discuss avenues to suppress the spin-to-charge conversion errors.

For initialization of Q2 in state $\ket{1}$, the dot should ideally remain empty at all times. Thus, imperfect spin-to-charge conversion arises either from
\begin{enumerate}
\item[1)] the finite time scales for an electron to leave or re-enter the dot. The conversion errors caused by these charge transitions could be suppressed by, e.g., engineering the density of states~\cite{mottonen2010} of the reservoir so that a spin-up electron tunnels out of the quantum dot very rapidly, and so that a spin-down electron tunnels back in the quantum dot very slowly~\cite{danjou2014}. If that were the case, the dot would remain empty at nearly all times and the bimodality of $P(I_p|1)$ would be suppressed.

\item[2)] spin relaxation of Q2 before the electron is able to leave the dot. Indeed, the spin relaxation time of Q2 is of order $1\,\textrm{ms}$, which is only an order of magnitude longer than the observed time scale $\sim 100\,\mu\textrm{s}$ for an electron to escape to the reservoir. The strategy to suppress spin relaxation of Q2 during readout depends strongly its physical origin. However, there is \emph{a priori} no fundamental reason why the spin relaxation time of Q2 cannot be as large as that of Q1 (although it must be noted that increasing the dot-reservoir coupling as suggested in point 1) may negatively affect the spin relaxation time via cotunneling processes).
\end{enumerate}

For initialization of Q2 in state $\ket{0}$, the dot should ideally remain occupied at all times. The probability of an electron leaving the dot should in principle be exponentially suppressed in the ratio of the thermal energy to the Zeeman splitting. However, the observed probability of transitioning to the reservoir is too large to be explained by such thermal suppression. Instead, a measurement of the transition probability as a function of the plunger gate voltage of Q2 suggests that the electron escapes the $\ket{0}$ state via an excited quantum dot state. The value of the plunger voltage at which the transitions are suppressed is consistent with an excited state with energy tens of $\mu\textrm{eVs}$ above the ground state. We conjecture that this state is an excited valley state with the same spin (spin-down) as the ground state. Therefore, a likely explanation for the dominant transition mechanism is excitation to a higher valley state (via, e.g., absorption of energy from the biased charge sensor) followed by a transition from the excited valley state to the reservoir. The bimodality of $P(I_p|0)$ could therefore be suppressed by engineering valley splitting much larger than the charge sensor bias.

The above discussion highlights the importance of understanding the underlying physics for optimization of qubit readout.

\bibliography{supplement}